\documentclass[12pt,english]{article}
\usepackage[T1]{fontenc}
\usepackage[latin9]{inputenc}
\usepackage[letterpaper]{geometry}
\usepackage{units}
\usepackage{amstext}
\usepackage{amssymb}
\usepackage{babel}

\begin{document}
\begin{center}
\textbf{From the Cosmological Constant: Higgs Boson, Dark Matter,
and Quantum Gravity Scales\medskip{}
}
\par\end{center}

\begin{center}
James R. Bogan 
\par\end{center}

\begin{center}
November 1, 2010\medskip{}
 
\par\end{center}

\textbf{Abstract} 

We suggest discovery targets for the Higgs boson, dark matter, and
quantum gravity mass scales, motivated by the Dirac equation for the
electron in deSitter space, and a sixth-order constraint between the
electron QED parameters and the cosmological constant. We go on to
show that this constraint can be viewed as a structural parameter
of the electron, and leads naturally to a new cosmic horizon. A dual
fourth-order constraint implies a second-order one, from which the
electron neutrino mass is derived.

\medskip{}

\textbf{I. Introduction} 

The Higgs boson really needs no introduction, as it is the focus of
world-wide attention by the particle physics community at FermiLab
and CERN, where the Large Hadron Collider (LHC) was built primarily
for its Higgs discovery potential. If found when the LHC begins analyzing
data from science runs, it will constitute the last missing link in
the highly successful standard model of particle physics. Since its
completion one-third of a century ago, the Glashow-Weinberg-Salam
(GSW) theory of electroweak (EW) interactions has rendered precise
predictions for the masses of the W and Z vector bosons {[}1{]}. In
contrast, the GSW theory can only assert that the Higgs mass is given
by its vacuum expectation value (\textasciitilde{} 246 Gev), scaled
by an unknown coupling constant, which is not computable from within
the SM {[}2{]}.

Outside the SM, versions such as the minimally supersymmetric standard
model (MSSM) have produced many predictions over the last decade {[}3{]}.
Indeed, some predictions of the Higgs mass lie outside the window
over which current experimental searches focus on, such as conformal
symmetry {[}4{]}, with $M_{H}$$\sim10^{-33}$ ev, to extra-dimensional
gauge fields {[}5{]}, with $M_{H}$ > 60 Gev, and to 5-D versions
of the SM {[}6{]} with $M_{H}>$600 Gev. The W and Z vector bosons
were subsequently observed in the early 1980's {[}7{]}. What has gone
unobserved, despite intense efforts at CERN and Fermilab is the Higgs
boson, the essential element to complete the GSW electroweak triad.
As of 1996, continuing attempts to refine the value of the W boson
and top quark masses has resulted in a shrinking of the theoretical
search window for the Higgs {[}8{]}, in concert with the experimental
one released last year, from the CDF/D0 collaboration at Fermilab
{[}9{]}.\smallskip{}

\textbf{Experimental Discovery Window} 

In March of 2007, it appeared that a dramatic narrowing of the search
window for the Higgs boson had arrived, when an indirect exclusion
window was published {[}10{]}, ranging from 114 - 144 Gev, at the
95\% confidence level (CL). Two years later, in March 2009, the upper
bound from the direct search window extended the range from the LEP2
lower bound of 114.4, up to and excluding the band from 160 to 170
Gev at 95\% CL {[}11{]}. Recent analysis has reduced this exclusion
band to 162 - 166 Gev {[}12{]}. Again in March of 2010, Erler {[}13{]}
has produced a powerful global analysis, culled from electro-weak
precision and Higgs search data which delineates a 90\% CL search
window for the Higgs boson, extending from 115 to 148 Gev, nearly
identical to that of Blazey {[}10{]}, as are seven predictions ranging
from 117 to 146 Gev from extra-dimensional theories {[}3{]}. The probablility
distribution in {[}13{]} is highly skewed toward the lower end, due
to prior LEP2 and Tevatron searches. We will see below that naturally
occurring scales point strongly to this region as well, favoring a
light Higgs.\newpage{}\smallskip{}

\textbf{Higgs boson-inflaton-electron connection}

In a recent paper {[}14{]}, Beck has presented a statistical argument
that there is an exact constraint between the properties of the electron
and the cosmological constant, anticipated earlier by Starobinsky
{[}51{]}. It can be expressed more transparently through the relation,
\smallskip{}

(1) $l_{p}^{2}\Lambda=\left(\frac{{\textstyle {\normalcolor l_{p}}}}{{\normalcolor {\textstyle \alpha\lambda_{e}}}}\right)^{6}$
,\smallskip{}

\begin{flushleft}
where $l_{p}$ is the Planck length, $\lambda_{e}$ is the reduced
electron Compton wavelength, $\alpha$ is the fine structure constant,
and $\Lambda$ is the cosmological constant, corresponding to a value
of the vacuum energy density which is in excellent agreement with
the WMAP value, namely 3.9 Gev/$m^{3}$, vs. 4.1 Gev/$m^{3}$ from
(1) {[}14{]}. This constraint is highly motivated by Dirac's 1935
equation {[}15{]} for the electron in deSitter space, written in modern
notation as,\medskip{}

\par\end{flushleft}

(2) $\gamma^{\mu}\gamma^{\nu}L_{\mu\nu}\psi=\sqrt{12/\lambda_{e}^{2}\Lambda}\,\psi$

\begin{flushleft}
where $L_{\mu\nu}=i(x_{\mu}\wedge\partial_{\nu})$ is the angular
momentum operator in the 5-dimensional deSitter space, and
\par\end{flushleft}

\begin{flushleft}
$\psi$ and $\gamma^{\mu}$ are the 5-dimensional Dirac spinors and
gamma-matrices respectively.
\par\end{flushleft}

\begin{flushleft}
As Dirac points out, this equation governs the quantum mechanical
motion of the electron on the 4-dimensional deSitter horizon. If we
presume spherical symmetry, then one can write (2) as an eigenvalue
equation,\medskip{}

\par\end{flushleft}

(3) $\sqrt{l(l+1)}\psi=\sqrt{12/\lambda_{e}^{2}\Lambda}\,\psi$ ;
\,solving for Lambda gives,\medskip{}
\medskip{}

(4) $l_{p}^{2}\Lambda=12\, l_{p}^{2}/\{\lambda_{e\,}^{2}l(l+1)\}$,\medskip{}

\begin{flushleft}
which indeed exhibits an inverse dependence of the cosmological constant
on the electron Compton wavelength. Equating (4) and (1) sets the
scale of the angular momentum quantum numbers, viz,\medskip{}

\par\end{flushleft}

(5) $l(l+1)=12\alpha^{6}\left(\frac{{\textstyle {\normalcolor \lambda_{e}}}}{{\normalcolor {\textstyle l_{p}}}}\right)^{4}\simeq5.8x10^{77};\, l\simeq7.6x10^{38}$,
as one might expect for a quantum mechanical system of cosmological
size.\medskip{}

The physical content of (5) is deep, since if one calculates the ratio
of the deSitter radius, $R_{dS}=\sqrt{3/\Lambda}$ to the zitterbewegung
radius $R_{z}=\left(\lambda_{e}/2\right)$, one obtains the value
computed in eq.(5) for \emph{l}, suggesting that the orbital angular
momentum of the Dirac electron on the deSitter horizon is related
to its spin, via the zitter motion {[}16{]}. Indeed, the Planck length
is most naturally interpreted as the geometric mean of the zitterbewegung
and Schwarzschild event horizon radii, implicit in the black hole
model of the Dirac electron {[}37-42{]}. Thus the deSitter and electron
event horizons should be interrelated. If we make the reasonable ansatz
that the number of electrons in the Dirac sea is proportional to the
product of their energy density and the volume of deSitter space,
we have,

\medskip{}

(6) $m_{e}c^{2}N_{e}=\rho V_{dS}$\,, and identify their energy density
with that of the dark energy, we find that\medskip{}

$N_{e}\simeq1.1x10^{83}=R_{dS}/R_{S}$.\medskip{}

\begin{flushleft}
This ratio of deSitter to Schwarschild radii has a compelling, geometrical
interpretation when inserted into (6) giving,\medskip{}

\par\end{flushleft}

(7) $m_{e}c^{2}=\rho_{cc}A_{dS}R_{S}=\frac{1}{3}\rho_{cc}V$ , where
V is the volume of a spherical shell formed by the deSitter horizon
area and the electron event horizon radius.

\begin{flushleft}
The energy resident in this macrosopic shell volume of $\simeq124\, cm^{3}$
corresponds to the rest mass of the electron.
\par\end{flushleft}

Taking the square root of (1), we can write it in terms of the the
cosmological constant mass as, \vspace*{\smallskipamount}

(8) $\left(\frac{{\textstyle m_{e}}}{{\textstyle \alpha\underline{m}_{p}}}\right)^{3}=8\pi\left(\frac{{\textstyle m_{cc}}}{{\textstyle \underline{m}_{p}}}\right)^{2}$,
where $\underline{m}_{p}=2.43x10^{18}$ Gev is the reduced Planck
mass. Thus (8) predicts,\medskip{}

(9) $m_{cc}$= 0.00237 ev\vspace*{\smallskipamount}

\begin{flushleft}
Indeed, there are several models in which the cosmological constant
is postulated to have a quantum electrodynamical (QED) origin {[}15-17{]}.
Although it is not understood whether the electron is the source of
the cosmological constant or vice versa, it is clear that (8) constitutes
a unique constraint between elementary particle and cosmological properties
which is not apparent in either standard model. Indeed for years now,
it has been part of the folklore that the proton bears a remarkable
relation to these parameters as well {[}20{]}, \smallskip{}

\par\end{flushleft}

(10) $\log\left(\frac{{\textstyle m}_{cc}}{{\textstyle M}_{p}}\right)=-11.6\simeq-\alpha^{-\frac{1}{2}}=-11.7$
. As a result, one now could argue that $m_{cc}(M_{p},m_{e},\alpha)$,

\begin{flushleft}
and is more naturally expressed in terms of just two dimensionless
parameters, $m_{cc}\left(\alpha,\mu\right)\,$, where $\mu$ is the
proton-electron mass ratio, $\simeq$1836.
\par\end{flushleft}

\begin{flushleft}
Together, dark energy and dark matter constitute $\simeq$ 96\% of
the energy density of the universe. As such, it seems unlikely that
either is uncorrelated with the other. Since the dark energy quantum
is traceable to QED, one wonders if dark matter might be as well.
Earlier this year, the CoGeNT experiment reported detection of a candidate
dark matter particle in the 7 - 11 Gev range {[}52{]}. Taking this
mass interval as a discovery target, we construct a dimensionless
hierarchy of dark matter to dark energy as approximately,\medskip{}

\par\end{flushleft}

(11) $\left(\frac{{\textstyle M}_{DM}}{{\textstyle M}_{CC}}\right)^{2}\simeq10^{25}$.

\begin{flushleft}
The other relevant hierarchy which is comparable in magnitude, is
the ratio of the Planck to electron masses, $\simeq2x10^{22}$. If
we form a ratio of these two scales, we obtain $\simeq500$. In his
large numbers hypothesis, Dirac insisted that such nearly identical,
large ratios of fundamental astrophysical parameters \emph{cannot}
be coincidental {[}53{]}. A particular example emerges from (17),
which when normalized by the electron mass-squared is $\simeq4x10^{42}$.
Dividing this number by the comparable ratio of the current age of
the universe to the electron Compton time scale, $\simeq3x10^{38}$,gives
a ratio of $\simeq12,000$. Thus, scale equality is achieved by a
factor of 24, or about $8\pi$, and we posit that,\medskip{}

\par\end{flushleft}

(12) $\left(\frac{{\textstyle M}_{DM}}{{\textstyle m}_{CC}}\right)^{2}=\left(\frac{{\textstyle \alpha}}{{\textstyle 8\pi}}\right)\left(\frac{{\textstyle m}_{p}}{{\textstyle m}_{e}}\right)^{3}\left(\frac{{\textstyle t}_{e}}{T_{U}}\right)$.\medskip{}

\begin{flushleft}
Using Teller's relation between the fine structure constant, Planck
time, and the age of the universe {[}54{]},
\par\end{flushleft}

(13) $T_{u}=8\pi t_{p}\exp(\nicefrac{1}{\alpha})=13.6\, Gy$ , we
obtain,\medskip{}

(14) $\left(\frac{{\textstyle M}_{DM}}{{\textstyle m}_{CC}}\right)^{2}=\alpha\exp(\nicefrac{-1}{\alpha})\left(\frac{{\textstyle \underline{m}}_{p}}{{\textstyle m}_{e}}\right)^{4}$,

\begin{flushleft}
which predicts a dark matter mass of ${\textstyle M}_{DM}=8.16\, Gev$,
in very close agreement with the recently obtained experimental value{[}55{]}.\medskip{}

\par\end{flushleft}

As Ozer points out {[}21{]}, there is a remarkable scale equivalence
between the Higgs, Inflaton, and, cosmological constant masses, \smallskip{}

(15) $\frac{{\textstyle m}_{cc}}{{\textstyle M}_{H}}\approx\frac{{\textstyle M}_{H}}{{\textstyle M}_{I}}\sim10^{-14}$\medskip{}

\begin{flushleft}
Since the vacuum energies associated to the epochs of Hubble acceleration
and inflation can be modeled by Higgs fields {[}22-23{]},we conjecture
that this proportion may correspond to an equality: \smallskip{}

\par\end{flushleft}

(16) $M_{H}^{2}=m_{cc}M_{I}$\medskip{}
\smallskip{}

Dimensional analysis suggests a natural, model-independent measure
of the inflaton mass. It has long been known that the electron charge
in combination with Newton's gravitational constant, defines an electro-gravitational
mass, which can be expressed in terms of the fine structure constant
and the Planck mass, viz,\smallskip{}

(17) $M_{EG}^{2}=\alpha\, m_{p}^{2}$\medskip{}

If we now construct a reduced electro-gravitational mass, which is
the geometric mean of the reduced Planck mass and a GUT-level mass
scale, such that, \smallskip{}

(18) $\underline{M}_{EG}^{2}=M_{GUT}\underline{\, m}_{p}=\alpha\,\underline{m}_{p}^{2}$;
such a mass scale is then given by,\medskip{}
\medskip{}

(19) $M_{GUT}=\alpha\,\underline{m}_{p}=1.78x10^{16}Gev$.\medskip{}
\medskip{}

This is in remarkable agreement with the MSSM-GUT scale of $2x10^{16}Gev$
{[}24{]}, as they are ostensibly identical. Taking the inflaton mass
as the geometric mean of this GUT-scale mass, and the mass corresponding
to the reheating temperature given in {[}22{]}, of $2.0x10^{15}Gev,$
we compute an inflaton mass of, \smallskip{}

(20) $M_{I}=\sqrt{M_{GUT}M_{reheat}}=6.0x10^{15}Gev.$\medskip{}

Inserting this value into (16) gives a Higgs mass of,\smallskip{}

(21) $M_{H}=120\, Gev$.\bigskip{}

We will see below, that the critical dependence of the Higgs mass
on the reheating temperature of inflation can be eliminated, yet the
value obtained is almost identical. In Erler {[}13{]}, the above value
for the Higgs mass is in close proximity to a narrow maximum in the
Higgs mass probability, peaked around 117 Gev, as is a recent computation
of the two-photon Higgs decay process {[}25{]}. These calculations
strongly suggest that a Higgs boson, with a mass near 120 Gev, may
be discovered in Tevatron or LHC data, and its value is traceable
to the constraint between the electron mass, the fine structure and
cosmological constants.\newpage{}

\begin{flushleft}
The implications for cosmological and particle physics are enormous,
since this interlinkage evolved from the electro-weak epoch to the
recent dark energy epoch, as a sort of `DNA' coding of our vacuum,
spanning 27 decades of time. It also explains why the value of the
cosmological constant is non-vanishing: all mass scales may ultimately
derive from it.\medskip{}

\par\end{flushleft}

\textbf{Electron-Neutrino Mass}

\medskip{}
If eq.(1) is rearranged slightly, a fourth-order constraint emerges,
viz,

(22) $R_{L}^{2}\Lambda=\left(\frac{{\textstyle {\normalcolor l_{p}}}}{{\normalcolor {\textstyle R_{L}}}}\right)^{4}$
, where $R_{L}=\alpha\lambda_{e}$ is the classical Lorentz radius
of the electron.

This suggests the existence of a second-order constraint, in terms
of new length and mass scales. It will be convenient to express the
former reduced by a factor of $\sqrt{8\pi}$, similar to the Planck
mass above as,\medskip{}

(23)\,$\frac{\lambda^{2}}{8\pi}\Lambda=\left(\frac{{\textstyle {\normalcolor l_{p}}}}{{\normalcolor {\textstyle R_{L}}}}\right)^{2}$
, where $\lambda$ corresponds to a macroscopic length scale of $2.46x10{}^{6}meters$,
and a microscopic mass scale given by,\medskip{}

(24) $m{}_{\lambda}c^{2}=8.02\, x10^{-14}ev$. \ It can easily be
shown that this mass scale derives from (8) in the form,\medskip{}

(25) $m{}_{\lambda}=\frac{1}{8\pi\underline{m}_{p}}\left(\frac{{\textstyle {\normalcolor m_{e}}}}{{\normalcolor {\textstyle \alpha}}}\right)^{2}$\medskip{}

Since the experiments on neutrino flavor-changing mandate a non-zero
rest mass for all three neutrino flavors {[}49{]}, one expects that
neutrinos would derive their mass from the Higgs field, and be scaled
in proportion to their associated lepton. Therefore we posit that
eq.(25), together with the Higgs mass, predict the electron-neutrino
mass to be given by,\medskip{}

(26) $m_{\nu_{e}}^{2}=\frac{M_{H}}{8\pi\underline{m}_{p}}\left(\frac{{\textstyle {\normalcolor m_{e}}}}{{\normalcolor {\textstyle \alpha}}}\right)^{2}\Rightarrow m_{\nu_{e}}=0.098$
ev.\medskip{}

\begin{flushleft}
Thus the electron neutrino possesses a rest mass of about one-tenth
an electron-volt, a value fully in accord with the particle data group
{[}50{]}, which lists an upper bound of 2 ev.
\par\end{flushleft}

\begin{flushleft}
Currently, the discovery bounds on the mass of the Higgs boson only
slightly constrain this prediction, as the 95\% confidence level ranges
from 115 to 158 Gev. This uncertainty reflects in the above value
as the square root of the Higgs mass, permitting the discovery window
for the electron-neutrino mass to vary from 0.094 ev to 0.129 ev.
It is intriguing to note that the above value of the electron-neutrino
mass in (26) is given almost exactly by,\medskip{}

\par\end{flushleft}

(27) $m_{\nu_{e}}=\alpha^{3}\frac{{\textstyle {\normalcolor m_{e}}}}{{\normalcolor {\textstyle 2}}}=0.099$
ev .\medskip{}

Equating (26) and (27), predicts a compelling value for the Higgs
mass, which is explicitly $\mathit{independent}$ of the electron
mass, viz,\medskip{}

(28) $M_{H}=2\pi\alpha^{8}\underline{m}_{p}=122.8$\,Gev.\medskip{}

Our previous estimate (21) was based upon the inflaton mass, computed
from the model-dependent reheating temperature (20). Equation (28)
requires no such reference to inflation, yet the value obtained is
only 2\% larger, implying a slightly higher inflaton mass \& reheating
temperature of:

\begin{flushleft}
$M_{I}=6.36x10^{15}Gev$ and $T_{RH}=2.27x10^{15}Gev$. \bigskip{}

\par\end{flushleft}

\textbf{II Tev-scale Quantum Gravity vs. Supersymmetry (SUSY)\medskip{}
}

Twelve years ago, an alternative resolution of the hierarchy problem
was announced {[}26{]}, in which large extra dimensions (LXD) effectively
lowered the scale of quantum gravity from the Planck scale of $10^{\text{19}}$Gev,
down to the Tev scale, accessible to particle accelerators such as
the Tevatron or the LHC, which might observe them directly. Thus a
reduced quantum gravity scale now vies with SUSY to resolve the hierarchy
problem. 

In a very recent paper {[}27{]}, Gasperini has convincingly argued
that the vacuum energy, first measured twelve years ago by astronomers,
owes its existence to SUSY breaking, which should manifest at the
Tev scale. A direct calculation of the upper bound gives, \smallskip{}

(29) $M_{SUSY}\leq\sqrt{m_{cc}\underline{m}_{p}}=2.4\, Tev$ \medskip{}

He goes on to argue that failure to observe SUSY effects, such as
sparticles below this scale, would likely constitute experimental
evidence against the existence of SUSY. This would leave the theory
of LXD as the primary candidate for solving the hierarchy problem.
Fortunately, a new test of LXD has just emerged {[}28{]}, which produces
a unique particle decay signature at a threshold near 6 Tev. Thus
we enquire as to whether the relations in (8) could predict an energy
scale consistent with this threshold. If there are six LXD, they constitute
a 6-volume, $V_{(6)}$ from which one can compute the reduced scale
of quantum gravitational effects as, \smallskip{}

(30) $M_{QG}^{4}=\sqrt{\frac{{\textstyle m}_{p}^{2}}{{\textstyle V}_{(6)}}}$\medskip{}

From differential geometry {[}29{]}, the volume of a 6-dimensional
hypersphere is given by,\medskip{}
\smallskip{}

(31) $V_{(6)}=\frac{{\textstyle \pi}^{3}}{{\textstyle 6}}$$R^{6}$\smallskip{}

Equations (1) \& (8) can be squared \& combined to give\medskip{}

(32) $\left(\frac{{\textstyle l_{p}}}{{\textstyle \alpha\lambda_{e}}}\right)$$^{6}=8\pi\left(\frac{{\textstyle m}_{cc}}{{\textstyle m}_{p}}\right)$$^{4}$,
Where $m_{p}$ is now identified as the Planck mass\smallskip{}

If we now identify the Lorentz radius of the electron (22) as the
radius of the 6-d hypersphere, we can express it in terms of the cosmological
constant mass as,\smallskip{}

(33) $R_{L}^{6}=\left(\frac{{\textstyle m}_{p}}{{\textstyle m}_{cc}}\right)^{4}\frac{{\textstyle l_{p}^{6}}}{{\textstyle 8\pi}}$\medskip{}
\smallskip{}

Thus from (31) we obtain the 6-volume,\medskip{}

(34) $V_{(6)}=\frac{{\textstyle \pi^{2}}}{{\textstyle 48}}$$\left(\frac{{\textstyle m_{p}}}{{\textstyle m}_{cc}}\right)^{4}$$l_{p}^{6}$\medskip{}
\smallskip{}

Inserting this into (30) gives, after some algebra,

(35) $M_{QG}^{2}=\left(\frac{{\textstyle 48}}{{\textstyle \pi}^{2}}\right)^{\frac{1}{4}}$$m_{cc}m_{p}$\medskip{}
\smallskip{}

Thus we arrive at a reduced quantum gravity scale given by,\smallskip{}

(36) $M_{QG}=6.55\, Tev$

Which is in excellent agreement with the threshold value cited in
{[}28{]}.

\medskip{}
\medskip{}

\textbf{III. The Dirac-Kerr-Newman black hole model of the electron\medskip{}
}

In this, the golden age of cosmology, one cannot help but reflect
back on the golden age of black hole research, 1963-1977, during which
most of today's current theory of the structure of classical black
holes was finalized. Today, this understanding is being applied and
extended to diverse areas of study, ranging from extreme black holes
in brane world theory, to billion-solar mass black holes powering
quasars. One would hope that similar pioneering efforts to understand
the nature of dark energy, both theoretically and experimentally,
will pay similar dividends in as little time. Twelve years after its
discovery, much progress has been made, and ambitious observational
projects are in the works to further refine other models, such as
quintessence {[}31{]} .

Yet it is fair to say that we are far from a universal consensus on
the nature and dynamics of the dark energy driving the acceleration
of the Hubble expansion. Most studies have employed astronomical methods,
with a continuing focus on supernovae {[}32,33{]},the ISW effect {[}34,35{]}
and the CMB {[}36{]}. These have reached perhaps the penultimate refinement
by the recent measurements of Vikhlinin et al.{[}37{]}, of the dynamics
of galactic clusters, in which three different measurement methodologies
have set the value of the equation of state parameter to the value:
w = - 0.991 +/- 0.045, which strongly suggests that a cosmological
constant (w = -1) affords the best fit to the accelerated Hubble expansion
data of the universe.

Working under this assumption, we show that the clearest physical
meaning of Beck's relation {[}14{]} is in the form of a cosmological
length scale relative to the Compton wavelength and Schwarzschild
radius of the electron. Adopting the Kerr-Newman model of the electron
{[}38-42{]}, as a spinning, charged black hole, we show that the modulus
of the complex event horizon is identical with the radius of the naked
ring singularity expressed in terms of these parameters, and that
a geometric mean of the Schwarzschild radius of the electron and a
new cosmological length scale is identical to the electron zitterbewegung
radius. We speculate about the interpretation of this new scale. 

Attempts to model the Dirac electron as a Kerr-Newman (DKN) black
hole were shown by Carter to yield exactly Dirac's value for the gyromagnetic
factor of g = 2 {[}36{]}. Ambitous attempts to further develop this
model have demonstrated a finite value for the self-energy {[}39{]},
charge and spin quantization {[}40{]}, a beautiful isometry between
the Kerr-Newman geometry and Dirac spinors {[}41,42{]}, and a prediction
that the positron may possess negative mass {[}43{]}. Nonetheless,
acceptance of the DKN model has been difficult, due to the absence
of detectable signatures in electron scattering experiments, Hawking
radiation, and the appearance of a naked singularity, brought out
most clearly by the expression for the event horizon radius, \medskip{}

(37) $R{}_{DKN}=$$\frac{R_{S}}{2}\left(1\pm\sqrt{1-\alpha\left(\frac{{\textstyle \lambda_{e}}}{{\textstyle l_{p}}}\right)^{2}-\frac{1}{4}\left(\frac{{\textstyle \lambda_{e}}}{{\textstyle l_{p}}}\right)^{4}}\right)$
, \medskip{}

where $R{}_{S}=1.35x10^{-57}m$ is the radius of the Schwarzschild
event horizon of the electron .\medskip{}

This is clearly complex, since the last term is on the order of $10^{89}$.
Using Beck's relation, (1), we can express (37) in terms of the cosmological
constant,\medskip{}

(38) $R_{DKN}=\frac{R_{S}}{2}\left(1\pm\sqrt{1-\left(\frac{{\textstyle 1}}{{\textstyle \alpha^{2}\lambda_{e}\sqrt{\Lambda}}}\right)-\left(\frac{{\textstyle 1}}{{\textstyle 2\alpha^{3}\lambda_{e}\sqrt{\Lambda}}}\right)^{2}}\right)$\medskip{}

If we compute the modulus of this complex event horizon radius, we
obtain the radius of the Kerr ring singularity itself,\medskip{}

(39) $R_{DKN}=\frac{{\textstyle R_{S}}}{{\textstyle 4\alpha^{3}\lambda_{e}\sqrt{\Lambda}}}=1.92x10^{-13}m=R_{z}$
,\medskip{}

Which is one-half the Compton wavelength of the electron, or Dirac's
\emph{zitterbewegung} radius. Thus it would appear that the cosmological
constant can be thought of as a structural parameter of the DKN electron-black
hole. The cancellation of factors of two in the expression for the
square of the Planck length,\medskip{}

(40) $l_{p}^{2}=R_{DKN}\times R_{S}=\left[\frac{\hslash}{2m_{e}c}\times\frac{2Gm_{e}}{c^{2}}\right]$,
\medskip{}

\begin{flushleft}
strongly suggests that black hole radii are the physical origin of
the quantum gravity scale, and not merely dimensionful groupings of
the fundamental constants, $\hbar$, G, and c. The algebraic structure
of eq.(39) is highly suggestive of another geometric mean, which we
write as,\medskip{}

\par\end{flushleft}

(41) $\lambda_{e}^{2}=R_{mv}\times R_{S}$ . A quick calculation yields,
\medskip{}

(42) $R_{mv}=1.10x10^{32}m=3.6$ Peta-parsecs (Ppc).\medskip{}

The hierarchy of current length scales in cosmology {[}44{]} is as
follows:\medskip{}

\begin{flushleft}
$R_{H}$ = 4.2 Gpc is called the Hubble scale for obvious reasons,
and is the length scale corresponding to our 13.7 Gy old universe.
\par\end{flushleft}

\begin{flushleft}
$R_{particle}$ = 14.1 Gpc is the more physically relevant particle
horizon, and corresponds to light signals emitted at the big bang.
\par\end{flushleft}

\begin{flushleft}
$R_{\infty}$ = 19.0 Gpc. We shall refer to $R_{\infty}$ as the eternal
horizon, and is the largest physical scale calculable in FRW cosmology.
\par\end{flushleft}

\begin{flushleft}
If our universe has an infinite lifetime, this distance is the farthest
we can ever directly observe. As such it is an information barrier
to events happening beyond it.
\par\end{flushleft}

\begin{flushleft}
$R_{mv}$ = 3.6 Ppc, exceeds both the particle and eternal horizon
scales of our universe \emph{by five orders of magnitude}.\medskip{}

\par\end{flushleft}

We conjecture that this may be the actual scale at which the presence
of the multiverse is manifest and predictable, as a spatial separation
between our eternal horizon, and that of a parallel universe. Indeed,
any speculation about the variation of the fine structure constant
over cosmological time scales would seem to be at odds with Beck's
relation. Thus it is quite plausible that any large variations in
a huge multiverse scale would correspond to small variations in the
cosmological constant, thereby rendering the fine structure term truly
constant, as is consistent with the evidence to date {[}56-57{]}.

One can only speculate how the cosmological constant influences the
properties of the electron, which if it can be interpreted as a Kerr-Newman
black hole, should possess only three good quantum numbers, namely
mass, spin, and charge to completely specify its properties. However,
it is clear that the electron charge and mass are degenerate with
respect to\,$\Lambda$, such that only $e$ \emph{or} $m{}_{e}$,
along with $\Lambda$ and spin are required. Indeed,\,$\Lambda$
appears to be electromagnetic in nature, despite the fact that dark
energy exerts a negative pressure, and would seem to preclude this.
Nonetheless, the proposal by Beck and collaborators {[}45{]} to search
for a $\Lambda$-driven cutoff in the noise spectrum of a Josephson
junction, would seem to depend critically upon an electro-gravitational
coupling of dark energy. 

\newpage{}

\textbf{Conclusion}

In the above heuristic formulae, we have calculated very plausible
values for the masses of the Higgs boson (122.8 Gev), dark matter
(8.16 Gev), LXD-quantum gravity (6.55 Tev), Inflaton ($6.36x10^{15}Gev)$,
and electron-neutrino (0.0993 ev). It appears that through its unique
linkage to the cosmological constant, the electron apportions scales
which probably correspond to black holes and the Higgs boson. Coincidentally,
it is also near the scale at which SUSY effects, such as sparticles,
are expected to manifest, if at all. It would be extraordinary in
defiance of Occam's razor, Nature accomodated \emph{two} such physical
phenomena capable of resolving the hierarchy problem. Hence one expects
to see one \emph{or} the other, but not both. Given the paucity of
experimental evidence for SUSY in three decades of searching, as well
as recent evidence against it, in which the measured B\_s meson switching
rate was much smaller than the SUSY prediction {[}30{]}, this author
feels that the odds favor observing a signature of quantum gravity
effects, such as enhanced lepton production beyond the SM prediction,
and/or black holes possibly later this year, now that the LHC has
begun operating at the 7.0 Tev center-of-mass energy level.

With regard to the conjectured multiverse scale $R_{mv}$, there is
no widespread agreement on any theoretical or experimental value.
After the recent observation of a large scale cosmic void {[}45{]},
speculation arose as to whether this void was a multiverse signature
{[}47{]}. Unfortunately, subsequent analysis of the data presented
compeling evidence that the void was simply an artifact of the statistics
used in its interpretation as such {[}48{]}. The multiverse scale
presented here is unexpected, in that it relies upon two measures
which one would think to be uniquely intrinsic to \emph{our} universe,
namely the cosmological and fine structure constants. Perhaps Nature
has made it possible to ascertain a scale for the multiverse in terms
of constants that nonetheless might vary from universe to universe.
If this is the case, it also follows that electrons possess a multiversal
reality via their description as Kerr-Newman black holes, and may
offer a microcosmic connection to the multiverse, complementary to
the astronomical one.

\bigskip{}

\textbf{References}

\begin{flushleft}
{[}1{]} `Introduction to the Standard Model of Particle Physics',
Cottingham \& Greenwood, Cambridge, 1998.
\par\end{flushleft}

\begin{flushleft}
{[}2{]} M.Gaillard, Rev.Mod.Physics, 71, \#2, S96-110, 1999
\par\end{flushleft}

\begin{flushleft}
{[}3{]} T.Schucker, arXiv: 0708.3344 {[}hep-ph{]}
\par\end{flushleft}

\begin{flushleft}
{[}4{]} R.Nesbet, arXiv:0811.4161 {[}hep-th{]}
\par\end{flushleft}

\begin{flushleft}
{[}5{]} Y.Hosotani, arXiv: 0809.2181 {[}hep-ph{]}
\par\end{flushleft}

\begin{flushleft}
{[}6{]} N.Haba, et.al., arXiv: 0910.3356 {[}hep-ph{]}
\par\end{flushleft}

\begin{flushleft}
{[}7{]} C.Rubbia, Rev.Mod.Phys., 57, \#3, 699-722
\par\end{flushleft}

\begin{flushleft}
{[}8{]} C.Quigg, arXiv: 0905.3187 {[}hep-ph{]}
\par\end{flushleft}

\begin{flushleft}
{[}9{]} K.Peters, arXiv: 0911.1469 {[}hep-ex{]}
\par\end{flushleft}

\begin{flushleft}
{[}10{]} CDF/D0: G.Blazey, APS mtg., 15 Apr. 2007
\par\end{flushleft}

\begin{flushleft}
{[}11{]} TEVNPH/CD-D0: arXiv: 0903.4001
\par\end{flushleft}

\begin{flushleft}
{[}12{]} CDF/D0: arXiv: 1001.4162 {[}hep-ex{]}
\par\end{flushleft}

\begin{flushleft}
{[}13{]} J.Erler, arXiv: 1002.1320 {[}hep-ph{]}
\par\end{flushleft}

\begin{flushleft}
{[}14{]} C.Beck, arXiv: 0810.0752 {[}gr-qc{]}
\par\end{flushleft}

\begin{flushleft}
{[}15{]} P.A.M. Dirac, Annals of Math., 36, 657-669, July, 1935.
\par\end{flushleft}

\begin{flushleft}
{[}16{]} D.Hestenes, Found. Physics., 20, 1213-1232, 1990
\par\end{flushleft}

\begin{flushleft}
{[}17{]} C.Beck, M.Mackey, Intl. J.Mod.Phys, D17:71-80, 2008
\par\end{flushleft}

\begin{flushleft}
{[}18{]} J.Beltran-Jimenez, Maroto, arXiv: 0811.0566
\par\end{flushleft}

\begin{flushleft}
{[}19 {]} Y.Fujii, K.Homma, arXiv: 0912.5164 {[}gr-qc{]}
\par\end{flushleft}

\begin{flushleft}
{[}20{]} F.Wilczek, personal comms., 2005
\par\end{flushleft}

\begin{flushleft}
{[}21{]} A.Ozer, arXiv: hep-ph/0408311
\par\end{flushleft}

\begin{flushleft}
{[}22{]} F.Bezrukov, M.Shaposhnikov, Phys.Letts. B569: 703-706, 2008
\par\end{flushleft}

\begin{flushleft}
{[}23{]} V. Dzhunushaliev, arXiv: 0907.5265 {[}gr-qc{]}
\par\end{flushleft}

\begin{flushleft}
{[}24{]} D.Lyth, A.Riotto, Phys.Rept., 314: 1-146, 1999
\par\end{flushleft}

\begin{flushleft}
{[}25{]} A.Arbuzov, et.al., arXiv: 0802.3427 {[}hep-ph{]}
\par\end{flushleft}

\begin{flushleft}
{[}26{]} Nima Arkani-Hamed, et.al., arXiv: hep-ph/9804398
\par\end{flushleft}

\begin{flushleft}
{[}27{]} M.Gasperini, arXiv: 0805.2330 {[}hep-th{]}
\par\end{flushleft}

\begin{flushleft}
{[}28{]} T.Plehn, D.Litim, Phys.Rev.Letts., 100, 131301, 2008
\par\end{flushleft}

\begin{flushleft}
{[}29{]} E.Weisstein. \textquotedbl{}Ball.\textquotedbl{} http://mathworld.wolfram.com/Ball.html 
\par\end{flushleft}

\begin{flushleft}
{[}30{]} CDF/D0; arXiv: hep-ex/0603029; Also see:
\par\end{flushleft}

\begin{flushleft}
http://www.fnal gov/pub/presspass/press\_releases/CDF\_meson.html
\par\end{flushleft}

\begin{flushleft}
{[}31{]} I. Zlatev et.al., arXiv: 9807002, 21Oct.1998
\par\end{flushleft}

\begin{flushleft}
{[}32{]} S.Perlmutter, et.al, Phys.Rev.Lett. 83 (1999) 670-673
\par\end{flushleft}

\begin{flushleft}
{[}33{]} Riess et.al.Astron.J.116:1009-1038,1998.
\par\end{flushleft}

\begin{flushleft}
{[}34{]} S.Boughn, R.Crittenden, Nature, 427, 45, 2004
\par\end{flushleft}

\begin{flushleft}
{[}35{]} R.Scranton et.al., arXiv:0307335
\par\end{flushleft}

\begin{flushleft}
{[}36{]} Y.Wang, P.Mukherjee, arXiv:0604051v2, 21Jun.2006
\par\end{flushleft}

\begin{flushleft}
{[}37{]} A. Vikhlinin et.al., arXiv: 0812.2720, ApJ, 692, 2009 Feb.10
\par\end{flushleft}

\begin{flushleft}
{[}38{]} B.Carter, Physical Review, 174, \#5,1559, 25 Oct.1968
\par\end{flushleft}

\begin{flushleft}
{[}39{]} S.Blinder, arXiv:0105029, 23 May 2001 
\par\end{flushleft}

\begin{flushleft}
{[}40{]} F.Cooperstock, V.Faraoni, arXiv: 0302080, 12 Feb.2003
\par\end{flushleft}

\begin{flushleft}
{[}41{]} H.Arcos \& J.Pereira, arXiv:0210103, 19Jan.2004; arXiv: 0710.0301
1Oct.2007
\par\end{flushleft}

\begin{flushleft}
{[}42{]} A.Burinskii, arXiv: 0612187, 9Jan.2007; arXiv: 0507109v4,
19Mar.2008
\par\end{flushleft}

\begin{flushleft}
{[}43{]} G. Chardin, Hyperne Interactions, 103, 83, 1997
\par\end{flushleft}

\begin{flushleft}
{[}44{]} M.Tegmark, Astrophys.J. 624, 463 (2005)
\par\end{flushleft}

\begin{flushleft}
{[}45{]} C.Beck, C.de Matos, arXiv:0709.2373.v2, 24Apr.2008
\par\end{flushleft}

\begin{flushleft}
{[}46{]} L.Rudnick, et.al., ApJ, 671, 40, 2007
\par\end{flushleft}

\begin{flushleft}
{[}47{]} M.Chown, New Scientist, 34-36, 24Nov.2007
\par\end{flushleft}

\begin{flushleft}
{[}48{]} L.Smith, D.Huterer, arXiv: 0805.2751, 18May 2008
\par\end{flushleft}

\begin{flushleft}
{[}49{]} M.Maggiore, `A Modern Introduction to QFT', Oxford U.Press,
2005
\par\end{flushleft}

\begin{flushleft}
{[}50{]} Particle Data Group; http://www.pdg.lbl.gov/
\par\end{flushleft}

\begin{flushleft}
{[}51{]} V.Sahni \& A.Starobinsky, Int.J.Mod.Phys.D9:373-444,2000
\par\end{flushleft}

\begin{flushleft}
{[}52{]} Nature news, 26 February 2010 | Nature | doi:10.1038/news.2010.97
\par\end{flushleft}

\begin{flushleft}
{[}53{]} P.A.M. Dirac, Nature, 139, p323, 1937
\par\end{flushleft}

\begin{flushleft}
{[}54{]} E.Teller, Physical Review, 73, \#7, 1 April, 801-802, 1948
\par\end{flushleft}

\begin{flushleft}
{[}55{]} D.Hooper, L.Goodenough, arXiv:1010.2752, 13 Oct.2010
\par\end{flushleft}

\begin{flushleft}
{[}56{]} J.Barrow, arXiv:0912.5510, 30 Dec.2009
\par\end{flushleft}

\begin{flushleft}
{[}57{]} S.Landau, C.Scoccola, arXiv: 1002.1603, 8Feb.2010
\par\end{flushleft}
\end{document}